\documentclass[manuscript,nonacm,screen]{acmart}

\usepackage{listings}
\usepackage{xcolor}
\usepackage[acronym]{glossaries}

\newacronym{dvc}{DVC}{Data Version Control}
\newacronym{dac}{DaC}{Data as Code}
\newacronym{fair}{FAIR}{findable, accessible, interoperable, and re-usable}
\newacronym{cli}{CLI}{command line interface}
\newacronym{iac}{IaC}{Infractsture as Code}
\newacronym{gui}{GUI}{graphical user interface}
\newacronym{dag}{DAG}{directed acyclic graph}
\newacronym{ads}{ADS}{automatic data serialisation}
\newacronym{mds}{MDS}{manual data serialisation}
\newacronym{ci}{CI}{continuous integration}
\newacronym{hpc}{HPC}{high-performance computing}
\newacronym{sdr}{SDR}{structural database repository}
\newacronym{rdf}{RDF}{radial distribution function}
\newacronym{md}{MD}{molecular dynamics}

\newcommand{\alias}[1]{%
  \ifcase#1%
    \or ZnTrack
    \or dask4dvc
    \or ZnDraw
    \or Package4
  \fi
}

\definecolor{codegreen}{rgb}{0,0.6,0}
\definecolor{codegray}{rgb}{0.5,0.5,0.5}
\definecolor{codepurple}{rgb}{0.58,0,0.82}
\definecolor{backcolour}{rgb}{0.95,0.95,0.92}

\lstdefinestyle{mystyle}{
    backgroundcolor=\color{backcolour},   
    commentstyle=\color{codegreen},
    keywordstyle=\color{magenta},
    numberstyle=\tiny\color{codegray},
    stringstyle=\color{codepurple},
    basicstyle=\ttfamily\footnotesize,
    breakatwhitespace=false,         
    breaklines=true,                 
    captionpos=b,                    
    keepspaces=true,                 
    numbers=left,                    
    numbersep=5pt,                  
    showspaces=false,                
    showstringspaces=false,
    showtabs=false,                  
    tabsize=2
}

\AtBeginDocument{%
  \providecommand\BibTeX{{%
    \normalfont B\kern-0.5em{\scshape i\kern-0.25em b}\kern-0.8em\TeX}}}

\begin{document}

\title{\alias{1}: Data as Code}

\author{Fabian Zills}
\affiliation{%
    \institution{Institute for Computational Physics}
    \streetaddress{Allmandring 3}
    \city{Stuttgart}
    \country{Germany}
    \postcode{70569}
}
\email{fzills@icp.uni-stuttgart.de}
\orcid{0000-0002-6936-4692}

\author{Moritz Schäfer}
\affiliation{%
    \institution{Institute for Theoretical Chemistry}
    \streetaddress{Pfaffenwaldring 55}
    \city{Stuttgart}
    \country{Germany}
    \postcode{70569}
}
\orcid{0000-0001-8474-5808}

\author{Samuel Tovey}
\affiliation{%
    \institution{Institute for Computational Physics}
    \streetaddress{Allmandring 3}
    \city{Stuttgart}
    \country{Germany}
    \postcode{70569}
}
\orcid{0000-0001-9537-8361}

\author{Johannes Kästner}
\affiliation{%
    \institution{Institute for Theoretical Chemistry}
    \streetaddress{Pfaffenwaldring 55}
    \city{Stuttgart}
    \country{Germany}
    \postcode{70569}
}
\orcid{0000-0001-6178-7669}

\author{Christian Holm}
\affiliation{%
    \institution{Institute for Computational Physics}
    \streetaddress{Allmandring 3}
    \city{Stuttgart}
    \country{Germany}
    \postcode{70569}
}
\orcid{0000-0003-2739-310X}

\renewcommand{\shortauthors}{Zills et al.}

\begin{abstract}
    The past decade has seen tremendous breakthroughs in computation and there is no indication that this will slow any time soon.
    Machine learning, large-scale computing resources, and increased industry focus have resulted in rising investments in computer-driven solutions for data management, simulations, and model generation.
    However, with this growth in computation has come an even larger expansion of data and with it, complexity in data storage, sharing, and tracking.
    In this work, we introduce \alias{1}, a Python-driven data versioning tool.
    \alias{1} builds upon established version control systems to provide a user-friendly and easy-to-use interface for tracking parameters in experiments, designing workflows, and storing and sharing data.
    From this ability to reduce large datasets to a simple Python script emerges the concept of \glsentrylong{dac}, a core component of the work presented here and an undoubtedly important concept as the age of computation continues to evolve.
    \alias{1} offers an open-source, FAIR data compatible Python package to enable users to harness these concepts of the future.
\end{abstract}






\maketitle

\section{Introduction}
\label{sec:introduction}
Large-scale computing is steadily becoming the norm in many industrial and research settings.
With the rise of large-scale compute clusters, computational sciences have risen equally as fast in the size and number of experiments being performed on the newly available resources~\cite{shalf20}.
This has coincided with extraordinary advances in machine learning which is now ubiquitous in science and industry applications.
Nevertheless, the challenge doesn't end with the capability to implement extensive simulations or machine learning models.
With the growing number of simulations and models comes an increase in parameters and workflows that must be tracked and stored efficiently, not only for reproducibility but also for distribution.
Furthermore, once a workflow is applied in production, it is often desirable to submit the results to a publication or, more often, share models with other users in one's community.
The challenges in managing this generated data remain an issue in many fields~\cite{allisonReproducibilityTragedyErrors2016, andersonIssuesBiomedicalResearch2007, pengReproducibleResearchComputational2011, stoddartThereReproducibilityCrisis2016, klumpVersioningDataMore2021}.
Sharing data in a standardized format can be challenging because there can either be many different formats to choose from or none that fulfill all the requirements of new research questions.
New formats can be introduced or generalizations attempted, but naming conventions and standardized data formats have a finite degree of complexity that they can handle before becoming overcrowded and unusable~\cite{gonzalezInteroperabilityNeutralData2007,oliveiraCECAMElectronicStructure2020}.
When discussing data, it is often overlooked that most of the time software was used to generate it in the first place.
This is to say that a simulation script has been written, a configuration file parsed, or \gls{cli} arguments have been used to initiate this generation of data.
In this case, it is reasonable to say that all information required to reproduce the data is contained in the code used in its generation.
Therefore, it is a convenient idea to provide a simple interface for sharing this code, along with its results, as data, i.e., to construct an interface capable of sharing \gls{dac}.

In this work, we present \alias{1}, a Python package designed to address these requirements.
The use of the Python programming language aligns with its popularity in the scientific community, especially in data science and machine learning~\cite{jetbrains2022survey}.
\alias{1} is built on the idea that once code has been written to generate data, no matter how complex the workflow, this is all that is required to reproduce or share the data.
Central to this goal is \alias{1}'s construction on top of the \Gls{dvc} framework, which provides a convenient interface to treat versions of code, i.e experiments, as commits or tags alongside potentially large data files inside a single repository driven by the GIT version control system.
\alias{1} allows users to store their data alongside the code used to generate it.
The usage of a data remote, together with well-established version control infrastructure such as GitHub, GitLab, or Bitbucket, is automatically managed by utilizing \gls{dvc}.
With \alias{1}, we combine the universal applicability of GIT and \Gls{dvc} with a dynamic and flexible interface driven by the Python programming language.

The purpose of this paper is to introduce the \alias{1} software to the community both on a technical level as well as through practical examples.
Initially, we introduce the concepts on which \alias{1} is built, these being computational workflow design on graphs, version control with GIT, and the \gls{dac} paradigm.
Following this overview of the theoretical aspects of the package, the architecture of \alias{1} is presented and discussed along with special mention of certain key features.
Finally, two use cases are explored in order to demonstrate the applicability and strengths of this new technology.
We showcase how \alias{1} can be used for purely Python-based workflows as well as how it can be expanded to work with other software.
\section{Related Work}
\subsection{Workflow and Data Management}
The concept of workflow and data management has been of interest for years.
Research groups and organizations have invested in the field on both theoretical and product-driven levels.
The result of this investment has been the emergence of general frameworks such as Apache Airflow~\cite{haines22}, kedro~\cite{Alam_Kedro_2023}, snakemake~\cite{molder21}, luigi~\cite{rieger17} and others~\cite{fitschen19, luo21, adorf18, george22, amstutz2022,daskdevelopmentteamDaskLibraryDynamic2016} which are widely used today. 
Most of these approaches target large and complex workflows and are for organizations comprising many researchers.
In most cases, a specialized setup process is required to interface with or construct a new form of these services.
Additionally, the majority of these pre-existing solutions depend heavily on databases hosted locally or on servers.
This introduces a significant degree of overhead for new users and typically requires experts in data structures and software engineering for maintenance.
Such problems might help to explain why many academic research groups have not adopted these strategies thus far~\cite{dasilvaCommunityRoadmapScientific2021,alamChallengesProvenanceScientific2022}.
Several groups have realized these shortcomings and developed tools to fill the needs left behind.
These include MLFlow~\cite{chen20} or wandb.ai~\cite{wandb} which are now widely known in the academic community and are even integrated into well-known software packages, for example, MACE~\cite{batatia22a} and NequIP~\cite{batzner22} from the quantum chemistry community.
In many cases, modern research software groups go to the trouble of integrating their own data management systems into their code as in pymatgen~\cite{ONG2013314}, MDSuite~\cite{tovey23a} or pyiron~\cite{JANSSEN201924}.
However, even in these cases, what is missing is a unified infrastructure for the storage and sharing of data.
To this end, products such as \Gls{dvc}~\cite{castro23} have emerged to combine parameter and workflow tracking with established version control tools like GIT.

\Gls{dvc} enables the use of version control tools like GIT for managing large data files.
Additionally, it provides a comprehensive workflow management system while remaining compatible with the entire GIT ecosystem.
\Gls{dvc} primarily utilizes a \gls{cli} and YAML configuration files, making it universally applicable to all file formats and the tools used to generate them.
However, this requires users to adapt their code to interface with \gls{dvc}, which leads to an overhead and might limit code flexibility.
To address some of these limitations, \gls{dvc} offers a Python API for accessing data, although it does not currently provide a public Python API for constructing or distributing workflows.

\subsection{Python Interfaces}
For the \gls{dac} concept, we want to take a look at existing datasets that can be accessed through Python packages.
This method of accessing datasets was already available in the early versions of TensorFlow~\cite{tensorflow2015-whitepaper} and PyTorch~\cite{paszkePyTorchImperativeStyle2019}.
An example of this is the MNIST~\cite{lecunGradientbasedLearningApplied1998} dataset, which is made available among other packages through  \verb|torchvision|  and can be downloaded through a simple API call.
\begin{lstlisting}[language=Python]
from torchvision import datasets
trainset = datasets.MNIST()
\end{lstlisting}
This concept is further improved by two of the most prominent tools for sharing training data or machine learning models, namely huggingface~\cite{wolfTransformersStateoftheArtNatural2020} and kaggle~\cite{KaggleYourMachine}, both of which provide a simple interface to download and use datasets.
\begin{lstlisting}[language=Python]
from datasets import load_dataset
dataset = load_dataset("mnist")
\end{lstlisting}
They provide not only a simple-to-use Python interface but also a sophisticated website that enables users to search for datasets hosted along with detailed descriptions of the available data, as well as different versions of the datasets if they are available.
Beneficial to these tools is the accessibility of metadata and the searchability provided by the website. 
However, their benefits are limited to a specific set of data made available through their platforms and they do not provide a general solution to all the challenges described above.

\subsection{Code-Driven Developments}
Datasets or models made available through code can be more easily integrated into workflows.
To fully benefit from the code interface, it is necessary to transition from static configuration files to dynamic script-based interfaces.
Due to its low barrier of entry and ubiquity in data sciences, the Python programming language is the ideal tool for this task.
One successful example of the transition from static configuration files to dynamic script-based interfaces is \gls{iac}~\cite{howardTerraformAutomatingInfrastructure2022}.
\Gls{iac} defines computing resources through code, rather than relying on specialized configuration files.
This approach provides greater flexibility and allows for the utilization of programming language features, such as loops and conditionals.
The scripts can be version-controlled in the same way as traditional configuration files, while being more transferable, allowing for better scaling, and typically being easier to maintain.
By automating parts that would be redundant in static configuration files, \gls{iac} also reduces the risk of misconfigurations.
Finally, the code can be documented in such a way as to make it easier to understand and maintain.
This flexibility is already available in workflow managers such as Airflow but is lacking in the data management tools mentioned above.
Bringing these concepts together can be seen as a prerequisite for \gls{dac}.

\section{Theory}
\label{sec:theory}
With this overview of related work, we want to introduce the \gls{dac} paradigm, showcase how it compares to existing solutions and highlight key differences.
Thereby putting the focus on workflow and data management as well as collaboration.

\subsection{Version Control with GIT}

Version control systems are essential tools in software development, allowing developers to track changes to source code and collaborate with other team members effectively.
GIT is a widely used distributed version control system that has gained popularity due to its efficiency, flexibility, and ease of use.
GIT uses a decentralized approach, allowing each developer to maintain their local repository and then merge changes with another repository when ready.
This approach provides several benefits, including faster processing times, improved collaboration, and the ability to work offline without requiring a connection to a central server.
One of the key features of GIT is its support for branching and merging.
Developers can create multiple branches of the same codebase, allowing for experimentation, feature development, and bug fixing without affecting the main branch.
GIT is compatible with various operating systems, making it easy for developers to work on different platforms.
Changes to the codebase are stored in GIT commits.
Each commit is assigned a unique identifier based on the changes made to the code.
This makes GIT ideal for human readable file formats.
On the contrary, large files, especially if they are compressed and a small change to the content will update the entire file, are unsuitable to be tracked with GIT.

\subsection{Data Version Control}
\gls{dvc}~\cite{castro23} can be employed to address the challenges posed by large data files in GIT repositories.

\begin{figure}[H]
    \centering
    \includegraphics[width=0.6\linewidth]{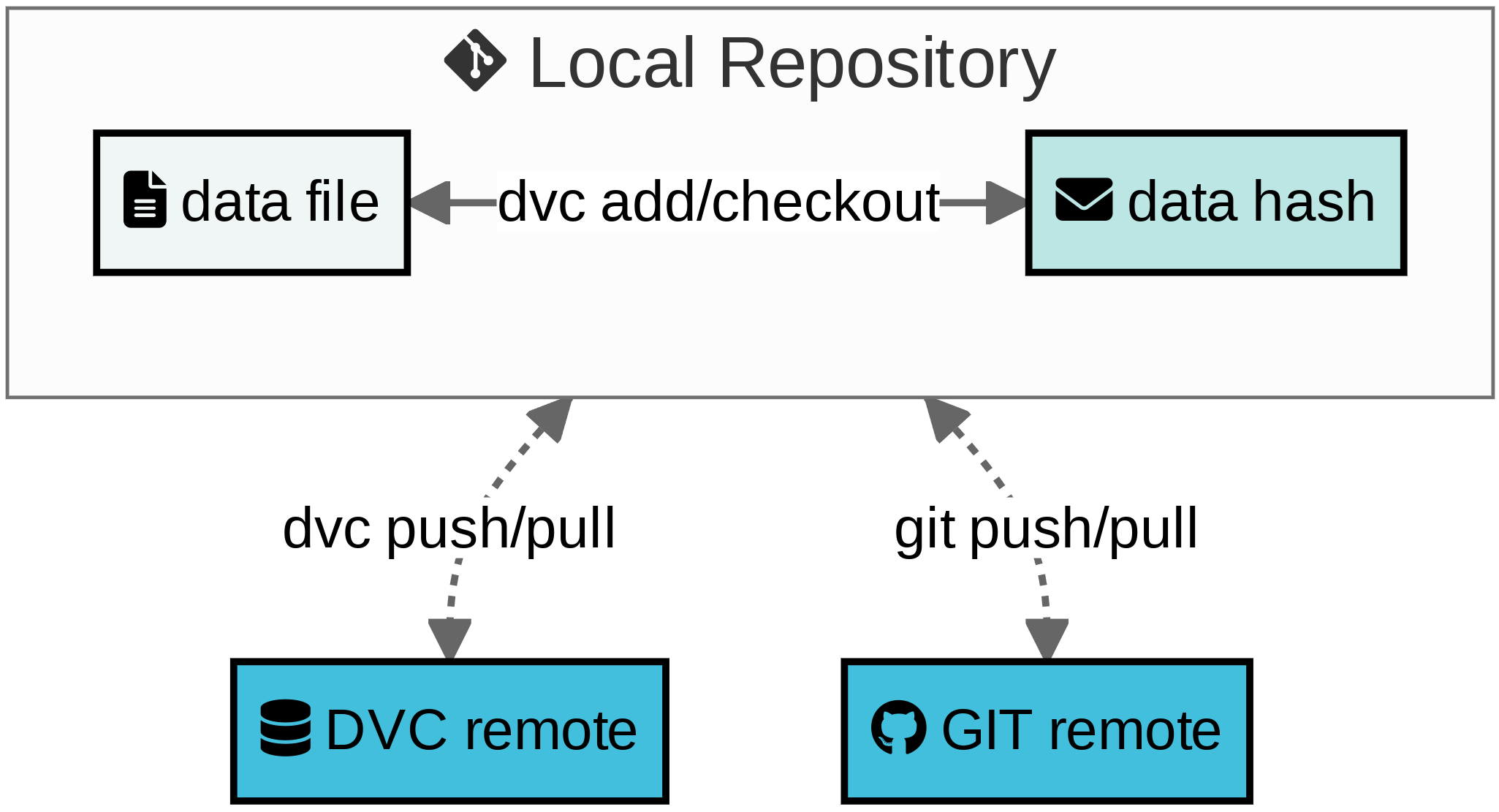}
    \caption{Combination of GIT and DVC in a local repository connected to a GIT remote and \gls{dvc} data remote.}
    \label{fig:git_dvc_concept}
\end{figure}

It achieves data versioning by computing a hash value for each file, and only this hash value is versioned using GIT.
When requested, the hash value can be utilized to retrieve the data associated with a specific commit from the data storage.
The storage options include local directories as well as remote storage such as object storage, WebDAV, and others (refer to Figure~\ref{fig:git_dvc_concept}).

Using GIT in conjunction with \gls{dvc} not only enables versioning of data but also promotes better collaboration and reproducibility.
These benefits make it evident why this combination is advantageous for research.









\begin{figure}[H]
    \centering
    \includegraphics[width=\linewidth]{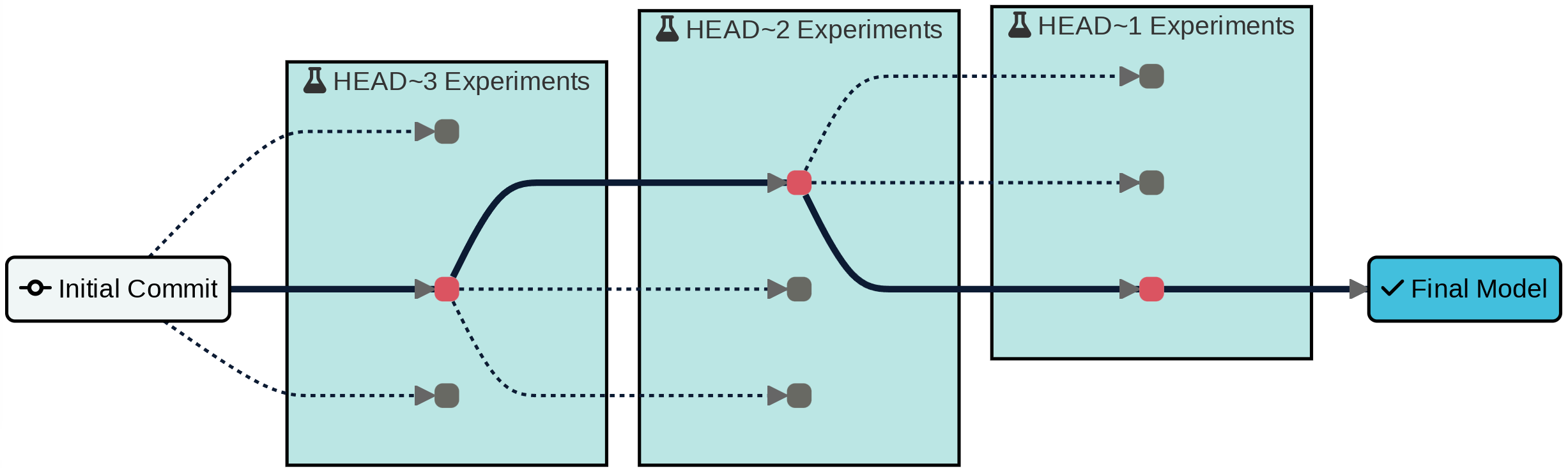}
    \caption{Experiment versioning using GIT. Each Experiment represents a detached commit. The best experiment is committed and new experiments are performed based on this commit.}
    \label{fig:git_exp}
\end{figure}

Version controlling both data and code in a GIT repository facilitates easy experimentation. By versioning input parameters, code, and outputs, experiments can be defined and stored as commits. 
Figure~\ref{fig:git_exp} illustrates how experiments can be performed in such a way.
The GIT and DVC \glspl{cli} enable comparison of experiments.
The best experiment can be promoted to a commit on a new branch, while changes to model architecture, training data, and parameters are tracked within a single repository.
This streamlined approach enhances the efficiency of experimentation and development processes.

Experiments can be compared using the \gls{dvc} \gls{cli} as well as \glspl{gui} such as the Visual Studio Code \gls{dvc} extension.
The corresponding data files can be accessed through a \verb|fsspec|~\cite{Filesystem_spec2023}-compliant Python API, that supports local and remote storage.

\subsection{Computational Graphs}
\label{sec:comp-graphs}


A core concept of modern computation is the representation of a workflow as a computational graph.
Within a computational graph, workflows are defined through nodes $N$ that represent variables or operations involved in a computation.
The flow of information and the dependencies between nodes are represented by the computational graph's edges $E$. Therefore, each connected node acts as a function on its inputs and passes its resulting outputs to its successors, as defined by the edges $E$~\cite{owhadiComputationalGraphCompletion2022,tinhoferComputationalGraphTheory2012}.
\begin{align}
    \begin{split}
        \text{Let } G &= (N, E) \text{ be a graph with a node set }\\
        N &= \{ 1, 2, 3, 4, 5 \} \text{ and an edge set }\\
        E &= \{ (1, 2), (1, 4), (2, 3), (4, 5), (5, 3) \}. 
    \end{split}
\end{align}
Each tuple in the edge set $E$ is a directed connection from one node to another.
An edge $(1, 2)$ from node 1 to node 2 indicates that the output of node 1 is used as an input to node 2. In other words, node 2 depends on the output of node 1.
\begin{figure}[H]
    \centering
    \includegraphics[width=0.5\linewidth]{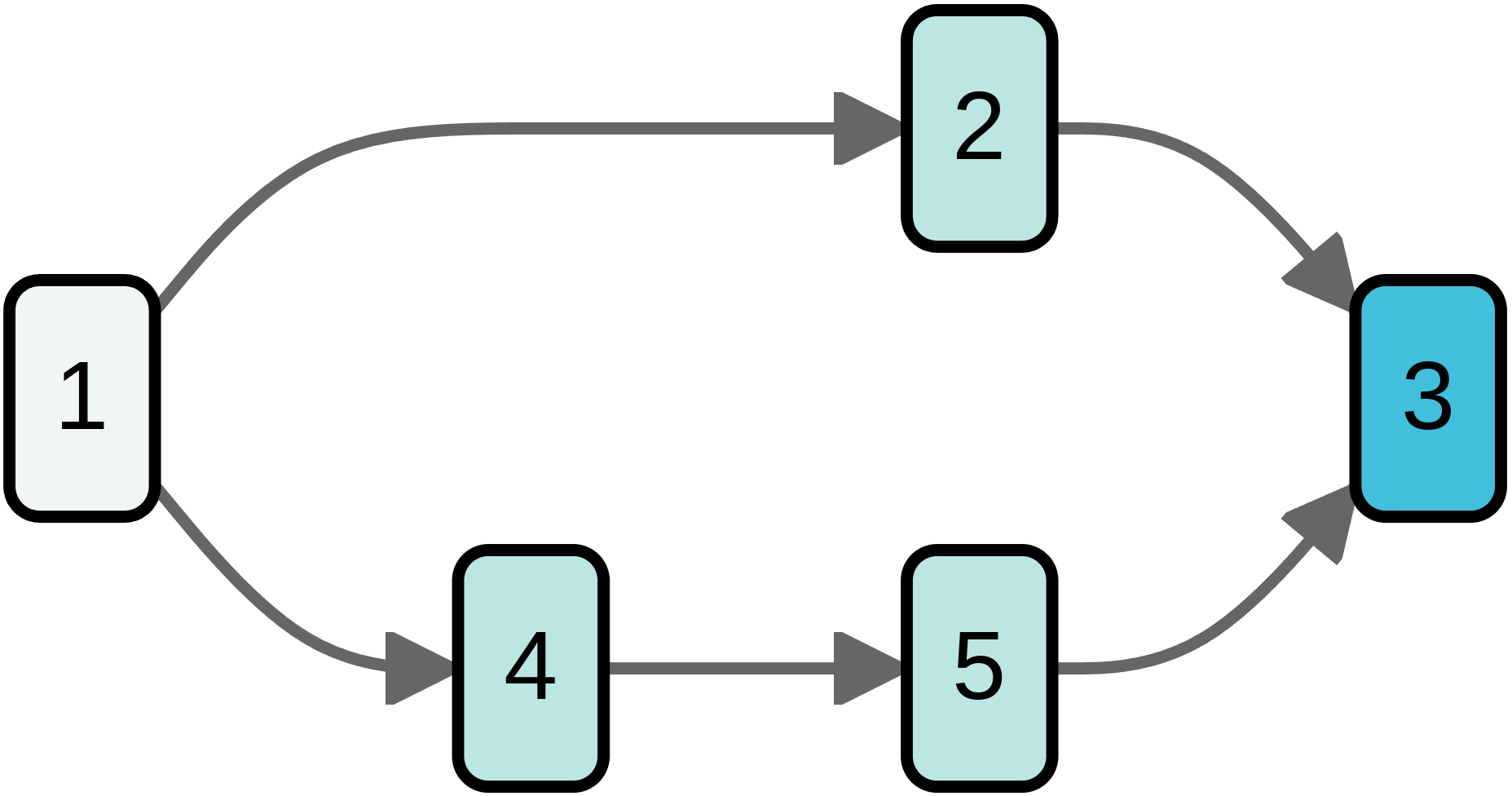}
    \caption{Illustration of a \gls{dag}.}
    \label{fig:dag}
\end{figure}
In the context of data-driven workflows, computational graphs are a powerful tool for managing dependencies, parameters and results.
We will focus on \glspl{dag} for the remainder of this work as illustrated in Figure~\ref{fig:dag}.
Connections within a \gls{dag} are directional, i.e. the information is only passed from the preceding node to the following one as indicated by the tuples in the edge set $E$.
Furthermore, it is prohibited to form cycles or loops~\cite{thulasiraman1992}.

Such graphs are broadly applied in machine learning and data analysis~\cite{lecunGradientbasedLearningApplied1998a,schulmanGradientEstimationUsing2015}.
Graphs allow for the identification of regions that are independent of each other and can be embarrassingly parallelized or optimized together~\cite{herlihyArtMultiprocessorProgramming2020,50530}.
They can be used in the context of automatic differentiation to compute gradients and are the basis of many machine learning frameworks such as TensorFlow~\cite{tensorflow2015-whitepaper}, PyTorch~\cite{paszkePyTorchImperativeStyle2019} or JAX~\cite{jax2018github}.

Computational graphs can be elegantly described in makefile-like formats, where the inputs, outputs, and corresponding functions are explicitly defined for each node.
Within the context of \gls{dvc}, a comprehensive toolset is provided to generate and persist these graphs in the YAML file format.
\Gls{dvc} leverages the concept of node-specific hash values, computed based on the inputs and outputs associated with each node.
This approach enables \gls{dvc} to efficiently determine when recomputation of results is necessary or if they can be loaded from a cache based on previous runs.

\subsection{Data as Code}
\begin{figure}[H]
    \centering
    \includegraphics[width=\linewidth]{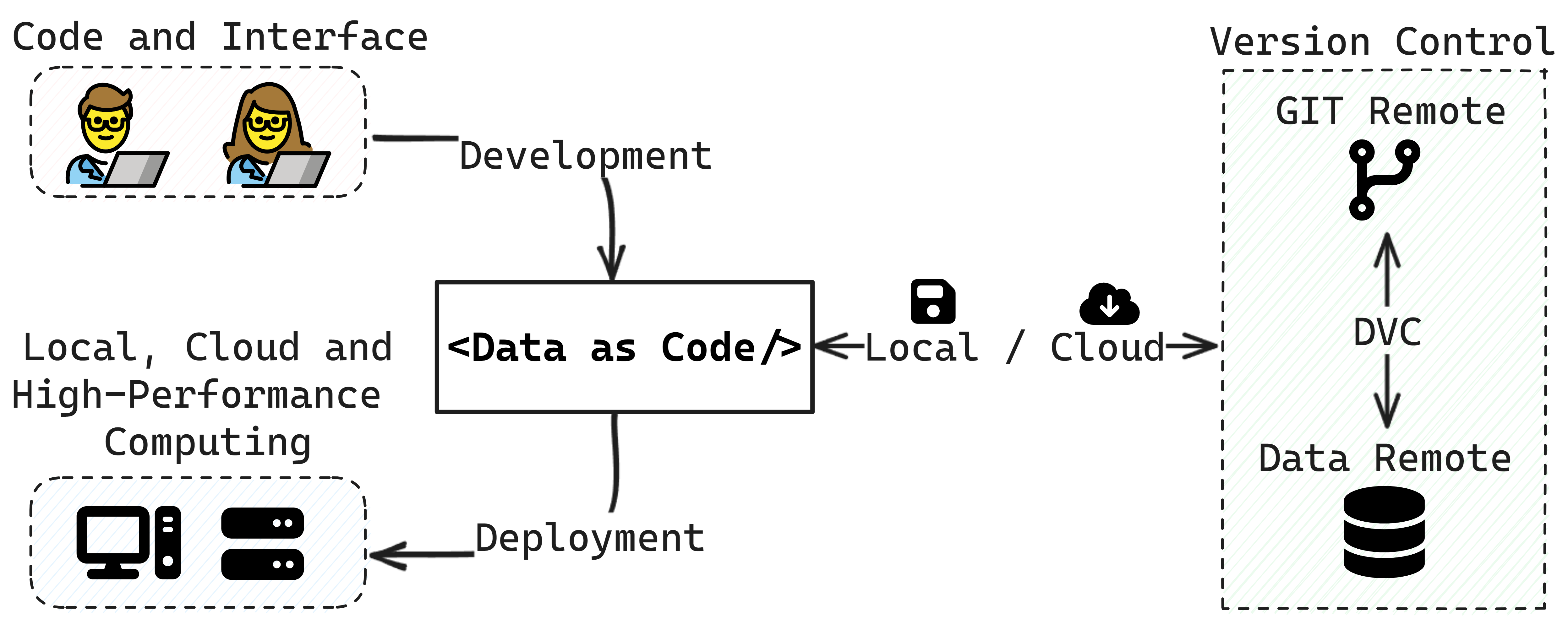}
    \caption{Illustration of the \gls{dac} paradigm.}
    \label{fig:dac}
\end{figure}

A key challenge in cooperating on data driven projects is the availability, accessibility and documentation of data sources.
Given a data repository that provides easy access to the data, a collaborator is still required to understand the chosen data format.
For making changes to the data, even more in depth knowledge about the construction of the data is required and essential parts of the computational workflow are often not shared alongside the data.
In the ideal scenario the data itself contains all of this information.
With \gls{dac}, we build on the idea that a single entity is responsible for the \textbf{generation}, \textbf{storage} and \textbf{interface} of data. 
It should be possible to version and share this entity with collaborators.
Such a framework is illustrated in Figure~\ref{fig:dac}.
The idea is to provide access to data by sharing the code that was used to generate it, thus rendering the problem of attached meta-data by an arbitrary format redundant as it is an integral component of the code itself.
By combining data generation and interfacing within the same entity, we ensure that the data remains readily accessible. 
Additionally, documentation is consolidated in a single location for both data generation and interface.
Nevertheless, compatibility can be ensured by abstraction of the interface and adherence to existing ontologies and code formats that are only altered if not sufficient for the given task.

In this code-centered data paradigm, multiple single-responsibility entities can be assembled to construct a computational graph, facilitating workflow management. 
These entities serve as the nodes on the computational graph. 
To enhance usability, when a user runs the shared code, cached files are loaded and made available instead of re-executing the data generation process. 
This approach promotes both workflow management and reproducibility, making them core features of the \gls{dac} concept.
Drawing inspiration from the transformative impact of version control on software development~\cite{brindescuHowCentralizedDistributed2014}, \gls{dac} provides a viable pathway towards achieving \gls{fair} data standards.

\subsection{FAIR Data and Code Style}
\label{sec:fair_solid}
For successful collaboration in data-driven research, it is important to agree on a set of principles that guide the development.
Introduced in 2016, the importance of \gls{fair} data~\cite{wilkinsonFAIRGuidingPrinciples2016} is more relevant than ever.
With the increasing amount of data that is generated, it is important to make data accessible and reusable.
In addition to the application of the \gls{dac} principles to achieve \gls{fair} data, good code practices are also essential.
Using an object-oriented approach, the SOLID principles are a good starting point for high-quality code design.
These principles recommend designing software in a way that is easy to maintain and extend~\cite{martin17}.
The SOLID principles refer to a set of guidelines, namely single responsibility~\cite{martinAgileSoftwareDevelopment2003}, open-closed~\cite{meyer1988object}, Liskov substitution~\cite{liskovKeynoteAddressData1987}, interface segregation~\cite{martinAgileSoftwareDevelopment2003} and dependency inversion~\cite{martinAgileSoftwareDevelopment2003}. 
In addition to these theoretical code design aspects, the use of opinionated code formatters and linters, along with detailed in-code documentation, can enhance collaboration.

\section{Architecture}
\label{sec:architecture}

In this section, we will introduce the architecture of the \alias{1} package.
We will highlight the concept of \gls{dac} and its implementation in \alias{1}.
Furthermore, we will describe the different features of the package in more detail and provide exemplary code snippets.

\subsection{\alias{1}}
To realize the \gls{dac} paradigm, data must be interfaced through code.
For this, an interface through a programming language is needed.
\Gls{dvc} and GIT, while useful for data and code versioning respectively, provide mostly \gls{cli} tools to generate a \gls{dac} infrastructure but do not fulfill all requirements to be considered \gls{dac} on their own.

\alias{1} builds on the synergy between GIT and \gls{dvc} to create a cohesive framework for establishing a unified repository where data and code coexist.
Embracing the \gls{dac} paradigm, \alias{1} empowers developers to effortlessly integrate data interfaces and code components in Python.
Through the definition of abstract dependency attributes and outputs, the code facilitates type checking and serves as a comprehensive documentation resource for both data and code. 

Additional benefits include the ability to easily share code and data with collaborators through platforms such as GitHub or GitLab. These platforms provide further infrastructure for managing issues, community discussions and code review.
Furthermore, whilst most current workflow management systems require a database, using GIT as the underlying framework allows for serverless and distributed workflow management.
This simplifies the migration of existing projects to a \gls{dac} workflow.
Users benefit doubly from this, as setup and maintenance efforts are drastically decreased.

The core components of the \alias{1} package are  split according to the components of a computational graph:
\begin{description}
    \item[Nodes] The \verb|Node| base class is the single interface to the code and the data. All data generation and data access are handled through this class. This includes not only the input data but also parameters, metrics, plots and the produced output data. Having a single interface responsible for one task strengthens the SOLID principles described in Section~\ref{sec:fair_solid}. Furthermore, it allows for documentation of the code and data in a single place. Data can be tested besides the code using common testing frameworks.
    \item[Edges] The \verb|Node| defines the expected dependency types but the connections between the \verb|Nodes| are handled by the \alias{1} \verb|Project|. The \verb|Project| acts as the interface to the computational graph. It handles the GIT and \Gls{dvc} commands and gives access to experiment handling. Through the \verb|Project| interface, \verb|Nodes| can also be organized in groups for a better overview.
\end{description}

To set up a \gls{dac} Project with \alias{1}, only the initialization of a repository with GIT and \Gls{dvc} is required.
This server-less setup makes it easy to create new projects with minimal overhead.
Later on, the projects can be pushed to a server to share the results with collaborators.

Structuring experiments in a very specific way can come with an overhead of time and might not be easily adapted by scientists from different areas.
To ensure a user-friendly introduction, \alias{1} provides all necessary tools to create a \gls{dac} infrastructure through a Python interface.
Minimal knowledge is sufficient to track your experiments using GIT and \Gls{dvc} in this way.
Additionally, \alias{1} enables the development of new tools that inherently incorporate \gls{fair} data management, parameter tracking, and distributed computing.





\subsection{Defining Nodes}
As described in Section~\ref{sec:theory}, a \verb|Node| is a single step in a computational workflow.
It is defined by a set of parameters and a function that is executed when the \verb|Node| is run.

At its core, a \verb|Node| can be described as a single function on the highest level control flow, i.e. what we from now on call the computational graph.
In the \alias{1} package, \verb|Nodes| can be defined as Python functions or classes.
In this work, our primary focus will be on the class-based approach.
The main advantage of using classes, as opposed to functions, is that they are stateful in this context, making it easier to access results.



By using classes to define \verb|Nodes|, we can store the parameters of the \verb|Node| as class attributes.
This capability allows us to conveniently access these parameters later on and establish a direct connection between the parameters, the data, and the function that is executed.

The Node class features an abstract \verb|run| method, which is intended to be overridden by the user. This allows the user to define the specific functionality that should be executed when the \verb|Node| is run.

\begin{figure}[H]
    \centering
    \includegraphics[width=0.8\linewidth]{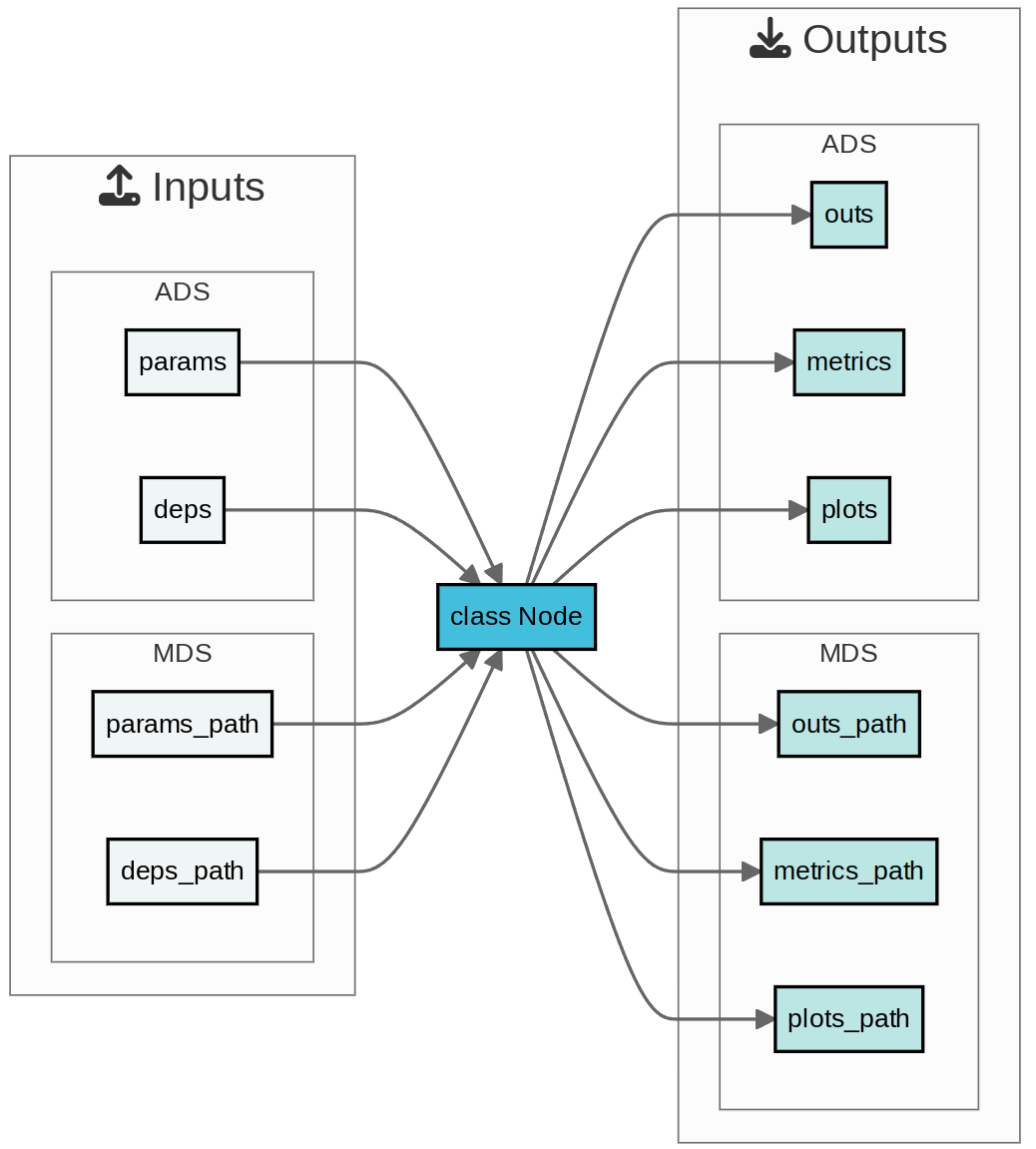}
    \caption{The inputs and outputs of a node are split into \gls{mds} and \gls{ads} attributes. The \gls{mds} attributes describe file paths whilst the \gls{ads} attributes can contain arbitrary data and are managed by \alias{1}.}
    \label{fig:track_options}
\end{figure}

To facilitate ease of use, parameters, dependencies, and outputs are defined as class attributes. These attributes are leveraged to automatically generate the class constructor. This automation simplifies the process of initializing and configuring Node instances.
It also allows for class inheritance and abstract classes, which can be used to define a common interface for a group of \verb|Nodes|.
The inputs and outputs to a \verb|Node| that are available in this way are summarized in Figure~\ref{fig:track_options}.
We differentiate between \gls{mds} and \gls{ads}.
Attributes suffixed with \verb|_path| are \gls{mds} attributes, i.e. they are file paths.
The way data is stored at these locations is entirely up to the user.
In contrast, attributes without the \verb|_path| suffix are \gls{ads} attributes.
These attributes are serialized by the \alias{1} package and provide a convenient way to store and access data.
Managing data from different commits or streaming data from the data remote is built into the \gls{ads} attributes, whilst for \gls{mds} attributes the user must make use of the \verb|Node.state.fs| filesystem interface.
The \verb|Node| can then be deserialized based on any commit in the repository.
Alongside the \verb|Node| all data from the specific commit will also be available.
With the availability of the code, a \verb|Node| can also be deserialized anywhere by supplying it with a link to the repository.
\begin{lstlisting}[language=Python]
import zntrack

class MyNode(zntrack.Node):
    data: str = zntrack.deps_path()
    result: float = zntrack.outs()
    shift: float = zntrack.params()

    def run(self) -> None:
        with open(self.data, "r") as f:
            self.result = float(f.read()) + self.shift
\end{lstlisting}












A simple example can be illustrated with the class \verb|MyNode|.
The \verb|Node| takes \verb|data| as an input file and adds a parameter called \verb|shift| to the \verb|data| value read from the file.
Finally, the \verb|Node| serializes the output variable \verb|result| of the computation to disk.
It is now possible to use \verb|MyNode| as a normal Python class and test its functionality.

\subsection{Defining Graphs}
To make use of \alias{1} and \Gls{dvc} features, we need to create a graph and add an instance of \verb|MyNode| to it.
Therefore, we instantiate a project, which manages the graph.
Within the project's context manager, connections between \verb|Nodes| are defined and parameters are set.
This code will only define the graph but not execute it.
All this information will be serialized and stored in human-readable configuration files that can be tracked via GIT.

\begin{lstlisting}[language=Python]
import zntrack

with zntrack.Project() as project:
    node1 = MyNode(data="data.txt", shift=1.0)
    node2 = MyNode2(value=node1.result)

project.build()
\end{lstlisting}

\subsection{Parallelization, Deployment and Interaction}
To execute the graph built within \verb|zntrack.Project|, multiple options are available.
All of these options share the common feature that a single command is all that's needed to assess the entire graph and execute only the necessary parts, i.e. only run the \verb|Nodes| with changed dependencies that haven't been executed yet.
This design essentially turns each \verb|Node| in the graph into a checkpoint, facilitating straightforward debugging and rerunning.

The \Gls{dvc} package only allows for sequential execution of the computational graph.
As described in Section~\ref{sec:comp-graphs} the computational graph can be used to parallelize the execution of \verb|Nodes| that are capable of running in parallel.
Therefore, we developed an additional graph executor that uses the Dask~\cite{daskdevelopmentteamDaskLibraryDynamic2016} package to parallelize the execution of the \Gls{dvc} workflow.
Our \alias{2} package enables the graph to be run in parallel and allows us to easily deploy the graph on a cluster, using the \textit{distributed} package. This allows for efficient utilization of computational resources when executing the graph.

Parallelization of a single graph is often not the only requirement for a scientific workflow.
In many scenarios, the same workflow must be executed with different parameters.
The concept of running the graph with different parameters is called an experiment in \Gls{dvc}.
In DVC, experiments are run using a queueing system based on \textit{Celery} and \textit{Hydra}.
\begin{lstlisting}
dvc exp run --queue -S MyNode.shift=1,2,3,4
dvc exp run --queue -S MyNode.shift=range(5, 10)
\end{lstlisting}
The code above shows how to queue experiments with different parameters.
If more flexibility is required, \alias{1} enables experimentation through Python, allowing for greater control and customization in the process.
\begin{lstlisting}[language=Python]
for x in range(5):
    with project.create_experiment() as exp:
        my_node.shift = x
\end{lstlisting}
The experiments can be executed from within Python or using one of the aforementioned graph executors.
This approach can also be combined with parameter optimization libraries such as Optuna~\cite{akiba2019optuna} or Ray Tune~\cite{liaw2018tune} to automate the process of finding optimal parameters whilst keeping all results and avoiding redundant computations.

Another powerful tool that can be brought to different branches of scientific research is the automatic deployment of predefined tasks such as model training or data analysis via \gls{ci} setups~\cite{cmldev}.
Changes in a parameter file can automatically trigger e.g. training of a new machine learning model using a \gls{ci} setup.
In this way, complex pipelines can technically be initiated with a click of a button even from mobile devices.
Another example would be the automatic analysis of experimental results through a \gls{ci} pipeline.
If a screening through many experiments with recurring analysis is required this can be automated and parallelized easily.

Finally, all of these steps are inherently fully reproducible, because all of the data is present in the repository or connected data storage.
This makes validating experiments much easier and fulfills the \gls{fair} principles whilst simplifying the process of storing experiment parameters and results in the same way.
This can even be achieved on a multitude of computing hardware, utilizing containerization on \verb|Nodes|.


\subsection{Analyzing Results}
A key task when running multiple experiments is the analysis and comparison of results.
There are multiple ways to achieve this.
The foremost is to use the \Gls{dvc} \gls{cli} to compare the results of different experiments.
This allows us to easily compare predefined metrics and plots.

\begin{lstlisting}
dvc metrics show
dvc metrics diff
\end{lstlisting}

A more convenient way to gather results is by loading the full \verb|Node| instance into a Python kernel.
This will make the parameters in combination with the results available through the \verb|Node| attributes.
To avoid loading not requested data into memory, all attributes are lazily evaluated.
This means that the data is only loaded when the attribute is accessed.

\begin{lstlisting}[language=python]
my_node.load()
\end{lstlisting}
If the instance is not present or multiple versions should be compared, one can also load the results from a specific revision and remote.
Possible revisions are the commit hash or a tag, as well as \Gls{dvc} experiment names.
\begin{lstlisting}[language=python]
my_node = zntrack.from_rev(
    "MyNode",
    rev="v2", 
    remote="https://github.com/user/repo"
)
\end{lstlisting}

With this approach, arbitrary analysis can be performed on the data.
Ultimately, an analysis node can be written with multiple nodes - even from different revisions or remotes - as dependencies and the analysis can be performed on the graph level.



\subsection{Collaboration}
Providing results to the scientific community and collaborating during the development of a project is also facilitated by the \gls{dac} paradigm.
Although all of the described principles can be used offline, their full potential is only realized when used in a collaborative environment.
\gls{dac} can be realized by sharing repositories on a platform like GitHub or GitLab.
Furthermore, specialized platforms such as Iterative Studio~\cite{incIterativeStudio} or DagsHub~\cite{OpenSourceData} allow more streamlined access to data and provide graphical user interfaces to run and compare results.
All of these tools are compatible with \alias{1}.

Additionally, \verb|Nodes| written with \alias{1} can be shared as Python packages and can be made available, e.g. through PyPi, building a \gls{fair} data and \gls{dac} infrastructure.
This allows others to access models and data easily.
As data and models are stored together, they are automatically compatible and can easily be used and adapted by others in their own projects.

This becomes even more important when considering that many new software applications will rely on data-driven methods and often require the inclusion of machine learning models.
These models can be too large to be effectively managed through GIT version control but still necessitate seamless integration into the codebase.
By adhering to \gls{dac} principles, one can avoid the need to version control the model and code separately, thus mitigating the risk of incompatibilities.


\section{Demonstration}
\label{sec:demonstration}
\subsection{Atomistic Simulation}
The \alias{1} package was developed with a strong focus on atomistic simulations.
These often require experimentation and can be computationally expensive.
Therefore, keeping track of all the experiments and their results is crucial.
In this section, we will demonstrate how \alias{1} can be used to manage and deploy computational graphs for such simulations.
Furthermore, we will introduce new ways of sharing simulation data and making results accessible to and reproducible by other researchers.
\subsubsection{Molecular Dynamics Simulation}
To illustrate the process, a \gls{md} simulation for a system of molten sodium chloride is deployed.
The example's workflow consists of generating a 3D structure from \verb|SMILES|~\cite{weiningerSMILESChemicalLanguage1988}, generating a simulation box and running the actual simulation.
We use the rdkit package~\cite{landrumRdkitRdkit20232023} together with packmol~\cite{martinezPACKMOLPackageBuilding2009} to generate the structure and \verb|Lammps|~\cite{LAMMPS} to run the simulation.




We run multiple experiments for simulation temperatures from 1000~K to 1800~K.
Each simulation contains 1000 ion pairs and is simulated for 100~ps.
Interactions are modeled by a Born-Meyer-Huggins-Tosi-Fumi (BMHTF) potential~\cite{pan16,fumi1964,huggins1933} and are analysed using ASE~\cite{larsenAtomicSimulationEnvironment2017}.
The experiments are queued using \verb|hydra| parameter composition and deployed in parallel using \alias{2}.
Using the \Gls{dvc} extension for Visual Studio Code, we can follow the progress of the experiments, e.g. the temperature, in real time.

\begin{lstlisting}
dvc exp run --queue -S "lammps.yaml:temperature=range(1000,1800,50)"
dask4dvc run --config config.yaml
\end{lstlisting}


\begin{figure}[H]
    \centering
    \includegraphics[width=0.7\linewidth]{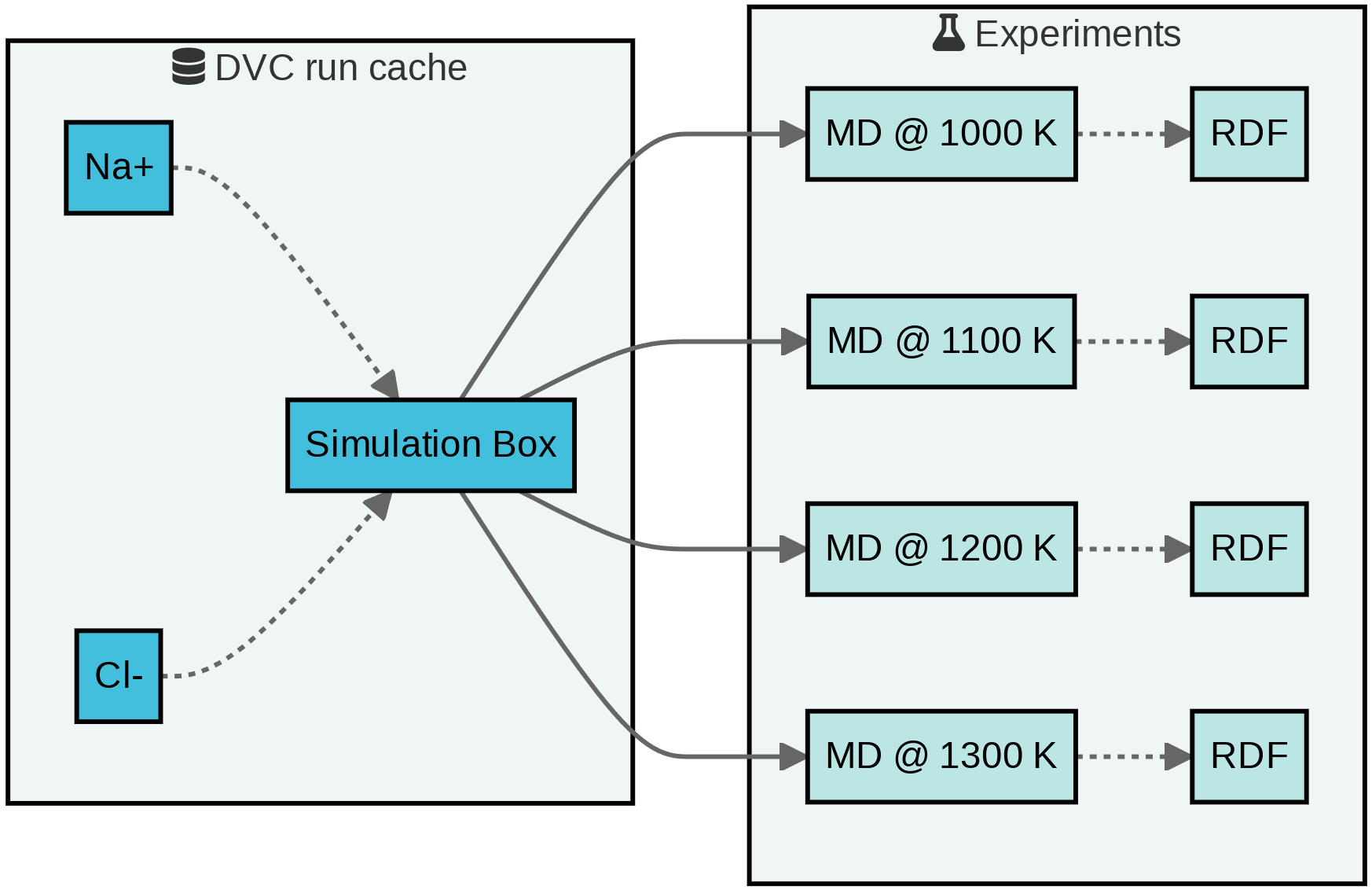}
    \caption{Parallel execution of multiple experiments.}
    \label{fig:dag_nacl_exp}
\end{figure}

After an initial structure has been generated, the respective nodes that generate the box are cached and the \gls{md} simulations are executed in parallel.
Figure~\ref{fig:dag_nacl_exp} visualizes the parts of the experiments that are only executed once and the parts that are executed for each experiment with different parameters.





\begin{figure}[H]
    \centering
    \includegraphics[width=\linewidth]{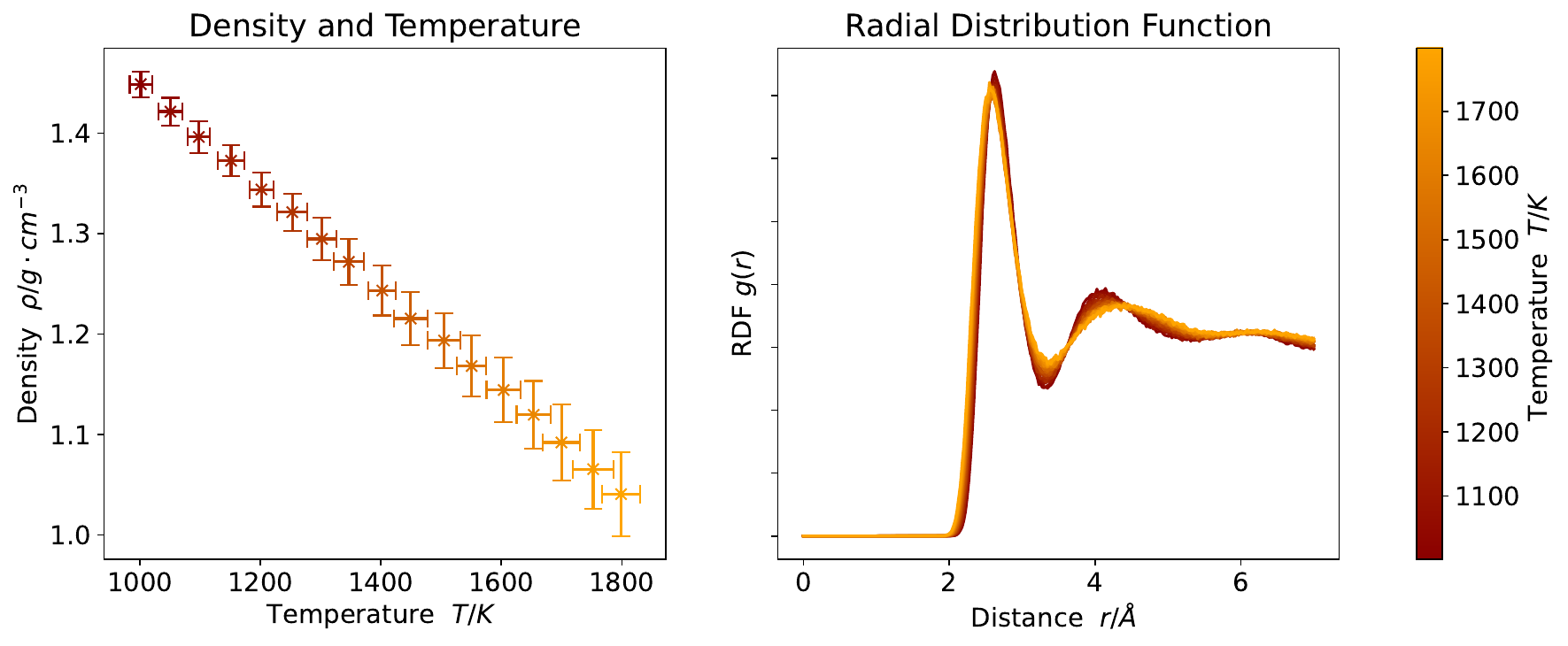}
    \caption{Density and \gls{rdf} of molten sodium chloride at different temperatures. The \glspl{rdf} are displayed for the Na-Cl pair.}
    \label{fig:nacl_analysis}
\end{figure}

For further comparison, the results of each experiment can be visualized using the \Gls{dvc} plot feature.
A more detailed analysis can be performed using the \alias{1} package.
Here, we can easily iterate over the experiments and load the results without making changes to the workspace.
Figure~\ref{fig:nacl_analysis} shows the density of the systems as well as the \gls{rdf} $g(r)$, in which temperature dependent structural changes can be seen.


Every experiment is referenced by a unique identifier and is associated with a commit.
All experiments are made available through the GIT repository and can be listed using \verb|dvc exp list <remote>|.
The \alias{1} package allows us to load the experiments directly. The only requirement for this to work is that the respective package for the \verb|Lammps| node is installed.
This node can then be inspected or connected to another computational graph.

\begin{lstlisting}[language=python]
node = zntrack.from_rev("Lammps", rev="exp1")
print(node.thermo)
\end{lstlisting}

\subsubsection{Structural Database Repository}
We have demonstrated how to use \alias{1} to run an atomistic simulation using Lammps and a BMHTF potential.
In recent years, machine learning has become a popular tool to model interatomic interactions instead.
These models are often trained on simulations from first principle methods.
Often, these datasets are shared in a way that either does not allow versioning them or will inflate the size of the repository by continuously editing large binary files that are not designed to be stored via GIT~\cite{eastmanSPICEDatasetDruglike2023,xiao2017/online}.
Many of these datasets are designed as benchmarks for machine learning models and therefore should not be altered. They are also used to train production models~\cite{ramakrishnan2014quantum,MachineLearningAccurate,smithANI1DataSet2017}.

With the \gls{sdr} we want to introduce a GIT repository with a public \Gls{dvc} remote that can be used to access, comment on and expand or improve first principle simulation data.
All information required to reproduce the data is available through this repository.
Data is stored in the H5MD format~\cite{debuylH5MDStructuredEfficient2014} and can be accessed either directly or through \alias{1} using ASE~\cite{larsenAtomicSimulationEnvironment2017}.
\begin{lstlisting}[language=python]
import zntrack

node = zntrack.from_rev(
    "BMIM_BF4_x10",
    remote="https://github.com/user/repo",
    rev="water"
)
node.atoms: list[ase.Atoms]
\end{lstlisting}

\begin{figure}[H]
    \centering
    \includegraphics[width=\linewidth]{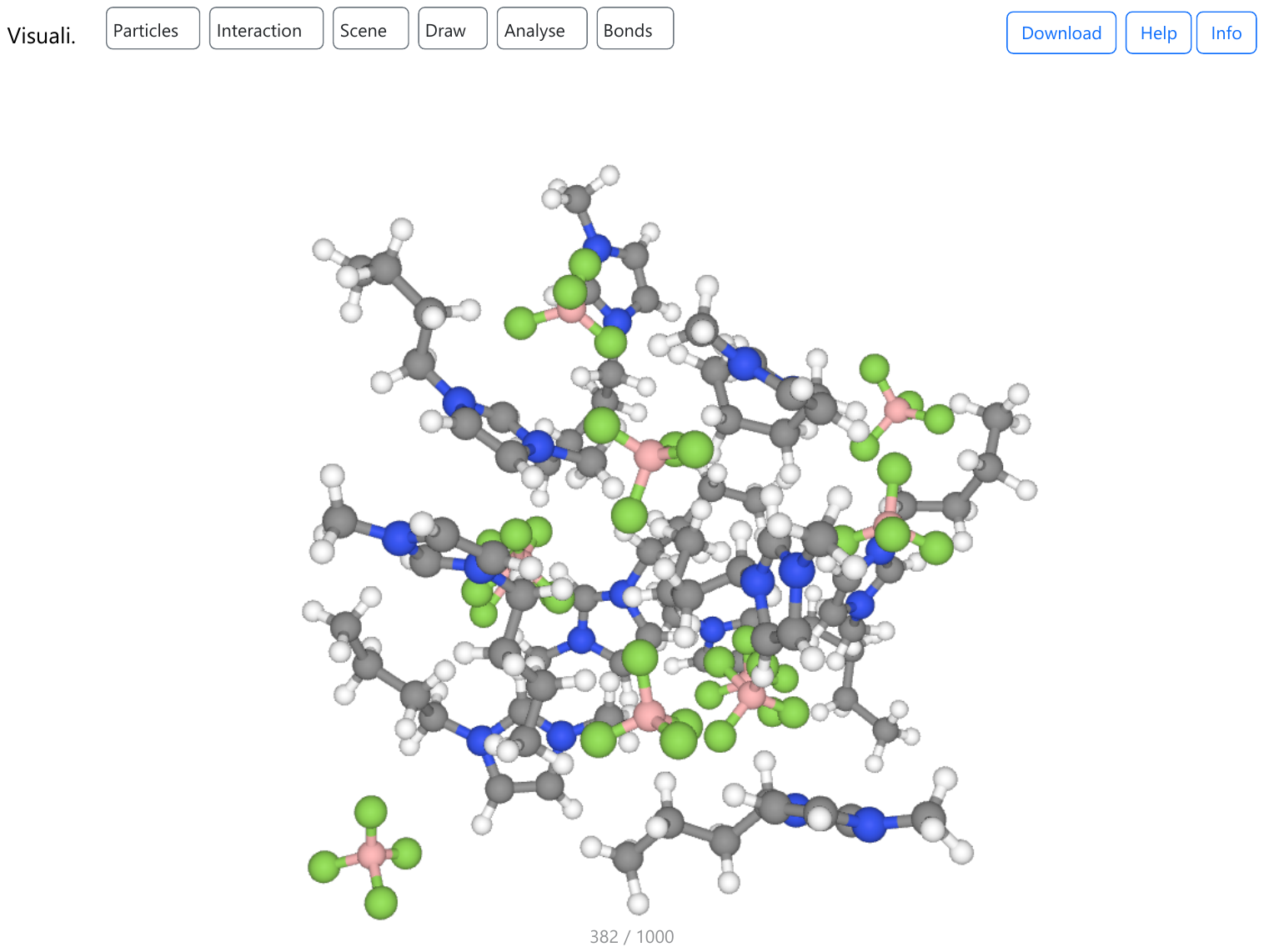}
    \caption{The room temperature ionic liquid BMIM-BF$_\textrm{4}$ visualized from the \gls{sdr} using \alias{3}.}
    \label{fig:bmim_sdr}
\end{figure}

Furthermore, tools can be built directly on top of \alias{1} to make working with the data even easier.
We provide the \alias{3} package that can visualize not only most molecular dynamics trajectory files but also access the \gls{sdr} directly as shown in Figure~\ref{fig:bmim_sdr}.

\begin{lstlisting}
zndraw BMIM_BF4_x10-x10.atoms --remote https://github.com/user/repo --rev BMIM-X
\end{lstlisting}

\subsection{Random Forest Classifier}
\alias{1} can also be used for more general machine learning applications.
We will showcase this by implementing a machine learning method for a binary classification task. 
To this end, we will demonstrate the training of a random forest classifier~\cite{breimanRandomForests2001} adapted from the \gls{dvc} example repository~\cite{IterativeExamplegetstartedGet} in \alias{1}.
This example utilizes a dataset consisting of StackOverflow questions and their respective tags.

\begin{figure}[H]
    \centering
    \includegraphics[width=0.9\linewidth]{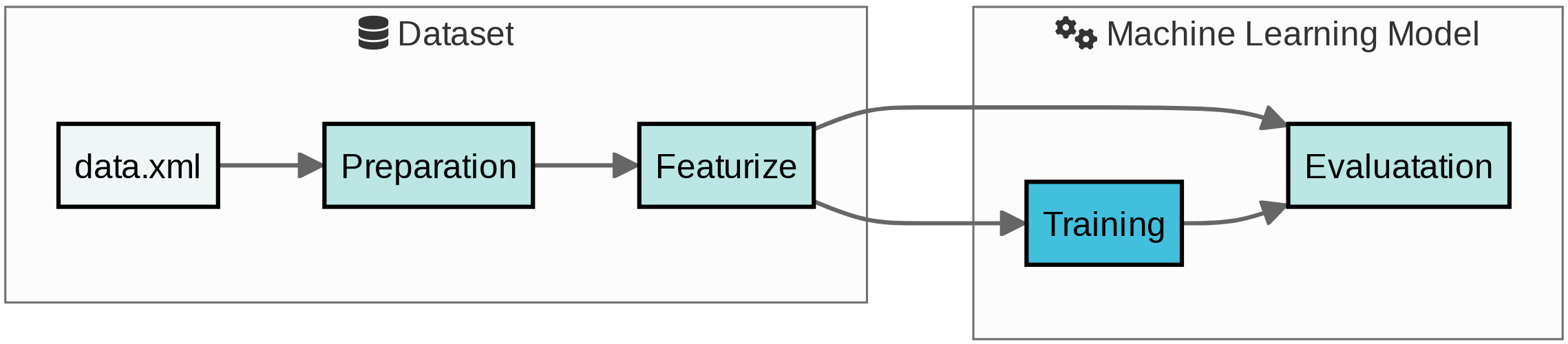}
    \caption{Computational graph for training a random forest classifier.}
    \label{fig:random_forest_dag}
\end{figure}
The code is organized into distinct steps, as illustrated in Figure~\ref{fig:random_forest_dag}.
The initial step involves extracting data from the raw data file and converting it into a suitable format.
Following this, the data undergoes featurization using a bag-of-words approach, and it is subsequently divided into separate training and test sets.
The features extracted from the training set are then employed to train a random forest classifier.
Lastly, the model's performance is evaluated using the unseen test set.
The code is implemented in Python and uses scikit-learn~\cite{scikit-learn}, numpy~\cite{harris2020array}, pandas~\cite{reback2020pandas} and matplotlib~\cite{Hunter:2007}.

\begin{figure}[H]
    \centering
    \includegraphics[width=0.9\linewidth]{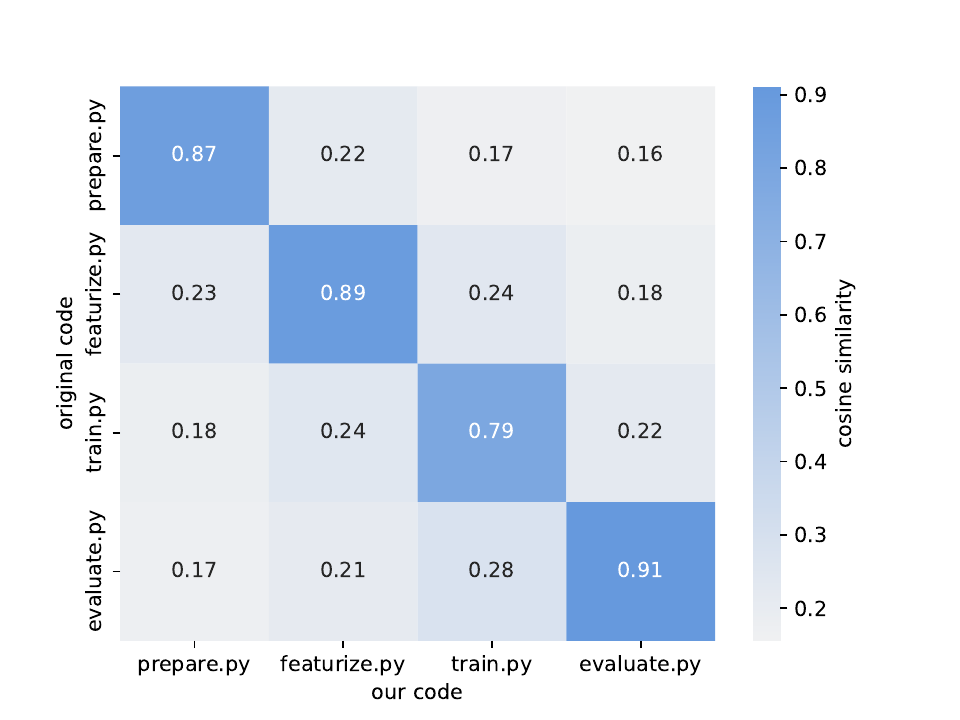}
    \caption{Similarity between the original code and the adapted code for \gls{dac} using \alias{1}. The similarity is calculated using the cosine similarity between the TF-IDF features~\cite{Manning:2008:IR} as implemented in scikit-learn~\cite{scikit-learn}.}
    \label{fig:random_forest_similarity}
\end{figure}

\begin{lstlisting}[language=python]
params = yaml.safe_load(open("params.yaml"))["train"]

inputs = sys.argv[1]
output = sys.argv[2]
seed = params["seed"]
n_est = params["n_est"]
min_split = params["min_split"]
\end{lstlisting}
We've converted the example code (see above) to adhere to the \gls{dac} paradigm by using \alias{1} (see code below).
For this scenario, we were able to reduce the lines of code and the McCabe complexity~\cite{McCabe1702388} between parts of the workflow with the same functionality.
Adapting data into the \gls{dac} paradigm through \alias{1} often only requires defining parameters, dependencies and outputs as class attributes, whilst the remaining code can be kept almost unchanged.
\begin{lstlisting}[language=python]
class Train(zntrack.Node):
    min_split: float = zntrack.params(0.01)
    n_est: int = zntrack.params(50)
    seed: int = zntrack.params(20170428)

    features: str = zntrack.deps_path("data/features")
    model: str = zntrack.outs_path("model.pkl")
\end{lstlisting}
This is further supported by the similarity analysis shown in Figure~\ref{fig:random_forest_similarity}.
In this analysis, we compute TF-IDF features~\cite{Manning:2008:IR} for both the original code and the code written with \alias{1}.
We then calculate the cosine similarity between them.
The results indicate that, despite adaptations, the new code remains similar to the original.
This strengthens the argument that \alias{1} can be used to convert existing code to adhere to the \gls{dac} paradigm, while keeping many of the original code's characteristics, thereby only requiring minimal changes and being not very invasive.



\subsection{Data Availability}
The described packages and conducted experiments are publicly available under the Apache-2.0 license.
\begin{itemize}
    \item \textbf{\alias{1}} \url{https://github.com/zincware/ZnTrack} and \texttt{pip install zntrack}.
    \item \textbf{\alias{2}} \url{https://github.com/zincware/dask4dvc} and \texttt{pip install dask4dvc}
    \item \textbf{\alias{3}} \url{https://github.com/zincware/ZnDraw} and \texttt{pip install zndraw}
    \item \textbf{Molecular Dynamics Simulations} \url{https://github.com/PythonFZ/zntrack_lammps}
    \item \textbf{Random Forest Classifier} \url{https://github.com/PythonFZ/DataAsCodeArtifacts}
\end{itemize}

\section{Conclusion}
\label{sec:conclusion}

In this work, we introduced the \gls{dac} concept, which simplifies access, sharing, and version control of data and workflows. 
\gls{dac} shifts the interface between data and code from end-users to developers and provides it alongside the code. 
We present the \alias{1} package, which facilitates \gls{dac} implementation in Python.

We demonstrated how DaC can be applied in atomistic simulations, allowing for easy parallel execution of multiple simulations and the FAIR sharing and version control of large datasets. 
Furthermore, we showed how packages can be built to interface with other \gls{dac} tools, as briefly demonstrated by the \alias{1} interface of the \alias{3} visualizer. 
Fully reproducible workflows support entry into data-driven fields, such as machine learning and atomistic simulations.

Leveraging DVC and GIT, \alias{1} offers easy integration of \gls{dac} into the research software and data lifecycle. It accommodates both newcomers and experienced researchers, providing minimal overhead for converting existing workflows. 

As an open-source package, \alias{1} also serves as a flexible workflow framework. It enables \gls{dac} within Python, promoting the close integration of code and associated data. This strengthens \gls{fair} data practices and fosters collaborations within and beyond research groups. Publishing research software, datasets, and experiments on platforms like GitHub or GitLab can contribute to the growth of open science, as seen in open-source research software.

\section*{Acknowledgements}
C.H., J.K., F.Z., and M.S. acknowledge support by the Deutsche Forschungsgemeinschaft (DFG, German Research Foundation) in the framework of the priority program SPP 2363, “Utilization and Development of Machine Learning for Molecular Applications - Molecular Machine Learning” Project No. 497249646.
S. T was supported by an LGF stipend of the state of Baden-W\"{u}rttemberg.
Further funding though the DFG under Germany's Excellence Strategy - EXC 2075 - 390740016 and the Stuttgart Center for Simulation Science (SimTech) was provided.
All authors acknowledge support by the state of Baden-Württemberg through bwHPC
and the German Research Foundation (DFG) through grant INST 35/1597-1 FUGG.

\bibliographystyle{ACM-Reference-Format}
\bibliography{bibliography.bib}

\end{document}